\documentclass[%
 aip,
 amsmath,amssymb,
 reprint,%
]{revtex4-1}

\usepackage{graphicx}% Include figure files
\usepackage{dcolumn}% Align table columns in on decimal point
\usepackage{bm}% bold math
\usepackage[utf8]{inputenc}
\usepackage[T1]{fontenc}
\usepackage{mathptmx}
\usepackage{etoolbox}
\usepackage{comment}
\usepackage{bm}
\usepackage[usenames]{color}
\usepackage{colortbl}
\usepackage[normalem]{ulem}
\newcommand{\red}{\textcolor{black}}

\makeatletter
\def\@email#1#2{%
 \endgroup
 \patchcmd{\titleblock@produce}
  {\frontmatter@RRAPformat}
  {\frontmatter@RRAPformat{\produce@RRAP{*#1\href{mailto:#2}{#2}}}\frontmatter@RRAPformat}
  {}{}
}%
\makeatother
\begin{document}

\preprint{AIP/123-QED}

\title[Mechanical scanning probe lithography of 2D perovskites for fabrication of planar polaritonic cavities]
{Mechanical scanning probe lithography of perovskites for fabrication of high-Q planar polaritonic cavities}
% Force line breaks with \\
\author{N. Glebov}
  \altaffiliation{These authors contributed equally}
  %\altaffiliation{nikita.glebov@metalab.ifmo.ru}%Lines break automatically or can be forced with \\

\author{M. Masharin}%
\altaffiliation{These authors contributed equally}%Lines break automatically or can be forced with \\

\affiliation{ 
ITMO University, School of Physics and Engineering, St. Petersburg, 197101, Russia%\\This line break forced with \textbackslash\textbackslash
}%

\author{B. Borodin}%
\affiliation{ 
Ioffe Institute, Saint-Petersburg 194021, Russia%\\This line break forced with \textbackslash\textbackslash
}%
\author{P. Alekseev}%
\affiliation{ 
Ioffe Institute, Saint-Petersburg 194021, Russia%\\This line break forced with \textbackslash\textbackslash
}%

\author{F. Benimetskiy}%
\affiliation{ 
Department of Physics and Astronomy, University of Sheffield, S3 7RH, Sheffield, UK%\\This line break forced with \textbackslash\textbackslash
}%
\author{S. Makarov}%
\affiliation{ 
ITMO University, School of Physics and Engineering, St. Petersburg, 197101, Russia%\\This line break forced with \textbackslash\textbackslash
}%
\affiliation{ 
Qingdao Innovation and Development Center, Harbin Engineering University, Qingdao 266000, Shandong, China}%
\author{A. Samusev*}
 %\homepage{anton.samusev@gmail.com}
 \email{anton.samusev@gmail.com}
\affiliation{ 
ITMO University, School of Physics and Engineering, St. Petersburg, 197101, Russia%\\This line break forced with \textbackslash\textbackslash
}%
\affiliation{%
Experimentelle Physik 2, Technische Universit\"at Dortmund, 44227 Dortmund, Germany%\\This line break forced% with \\
}%

\date{\today}% It is always \today, today,

\begin{abstract}
% An article usually includes an abstract, a concise summary of the work covered at length in the main body of the article. It is used for secondary publications and for information retrieval purposes. 

Exciton-polaritons are unique quasiparticles with hybrid properties of an exciton and a photon, opening ways to realize ultrafast strongly nonlinear systems and inversion-free lasers based on Bose-Einstein polariton condensation. However, the real-world applications of the polariton systems are still limited due to the temperature operation and costly fabrication techniques for both exciton materials and photon cavities. 2D perovskites represent one of the most prospective platforms for the realization of strong light-matter coupling since they support room-temperature exciton states with large oscillator strength and can simultaneously be used for the fabrication of planar photon cavities with high field localization due to the high refractive index of the material. In this work, we demonstrate  the \red{affordable for research purposes} mechanical scanning probe lithography method for the realization of room-temperature exciton-polariton systems based on the 2D perovskite (PEA)$_2$PbI$_4$ \red{with the Rabi splitting exceeding 200~meV}. By the precise control of the lithography parameters, we broadly adjust the exciton-polariton dispersion \red{and, in particular, vary the radiative coupling of polaritonic modes to the free space}. Our findings represent a versatile approach to the fabrication of planar high-quality perovskite-based photonic cavities supporting the strong light-matter coupling regime for the development of on-chip all-optical active and nonlinear polaritonic devices.

\end{abstract}

\maketitle

%\begin{quotation}
% The ``lead paragraph'' is encapsulated with the %\LaTeX\ 
%\verb+quotation+ environment and is formatted as a %single paragraph before the first section heading. 
%(The \verb+quotation+ environment reverts to its usual meaning after the first sectioning command.) 
%Note that numbered references are allowed in the lead paragraph.
%
%The lead paragraph will only be found in an article being prepared for the journal \textit{Chaos}.
%\end{quotation}

%\section{\label{sec:level1}Introduction} 
%\protect\\ The line
%break was forced \lowercase{via} \textbackslash\textbackslash}

Photonics deals both with fundamental and applied aspects of operating with optical signals, as well as with prospective designing energy-efficient optical computing devices. Implementing such devices where light is controlled by light requires systems with strong optical nonlinearity. Optical systems with the strong coupling of photon cavity mode with an exciton resonance, resulting in exciton-polariton, demonstrate a nonlinear response up to 3-4 orders of magnitude higher than in weakly coupled systems.\cite{belanger1997solitonlike} Such systems are realized by embedding an excitonic material with high exciton oscillator strength into a photon cavity supporting a mode with strong field enhancement and a long radiative lifetime. \cite{kavokin2017microcavities} The search for excitonic materials as well as the design of photon cavities suitable for the incorporation with efficient fabrication methods is therefore of great importance for polaritonics today.

One of the most studied and widely used material platforms for the exciton-polariton systems is the GaAs quantum well (QW), embedded into the vertical Bragg cavity.\cite{weisbuch1992observation} Due to the low exciton binding energy, the operation of these polariton systems is limited to cryogenic temperatures.\cite{bajoni2008polariton,kasprzak2006bose} The temperature limitations can be overcome with wide-gap semiconductor QWs such as ZnO\cite{van2006exciton} or GaN \cite{semond2005strong}, but they still require time-consuming and costly fabrication methods such as epitaxial growth techniques. Monolayer transition metal dichalcogenides have become perspective materials for room-temperature polariton systems, \cite{stepanov2021exciton,kravtsov2020nonlinear} though their potential applications are still limited by technological scalability. Currently, halide perovskites represent the promising \textcolor{black}{platform} for exciton-polariton systems due to their easy and cost-efficient fabrication as well as their outstanding excitonic properties making it possible to implement room-temperature exciton-polariton systems.\cite{su2021perovskite} Moreover, two-dimensional perovskites with enormous exciton binding energy \textcolor{black}{in the range of $190-400$~meV\cite{gelvez2020overcoming}} and exceptionally strong excitonic response \cite{zhao2018layer} have experimentally demonstrated the record-high value of Rabi splitting among perovskites exceeding 200~meV at room temperature \cite{wang2018room} and therefore represent one of the promising materials for polaritonic systems.

The most commonly used photon resonator in polaritonic systems is the vertical Bragg cavity since it provides all necessary requirements such as low optical losses, controllable lifetimes, and high field enhancement.\cite{butte2005recent} Exciton-polaritons in perovskite materials and also in 2D-perovskites have been already demonstrated in the Bragg resonators.\cite{wang2018room,su2020observation,wu2021nonlinear} Nevertheless, such structures \red{are bulky, i.e.} have large vertical sizes and also require sophisticated and costly fabrication methods, which severely hinder real-world applications.\cite{sturm2009strong} Meanwhile, compatible with on-chip designs planar photon cavities, such as metasurfaces or photonic crystal slabs (PCSs) can demonstrate comparable characteristics and have been recently employed in exciton-polariton systems with various materials.\cite{whittaker2021exciton,kravtsov2020nonlinear,masharin2022polaron} Moreover, high-Q symmetry-protected bound states in the continuum (BICs), appearing in metasurfaces, when strongly coupled to the exciton resonance,\cite{koshelev2018asymmetric} allow to even realize polariton Bose–Einstein condensation.\cite{ardizzone2022polariton} Although planar photon cavities based on perovskites are more suitable for future applications, there is still a lack of efficient and low-cost cavity fabrication techniques.

\begin{figure}
\includegraphics{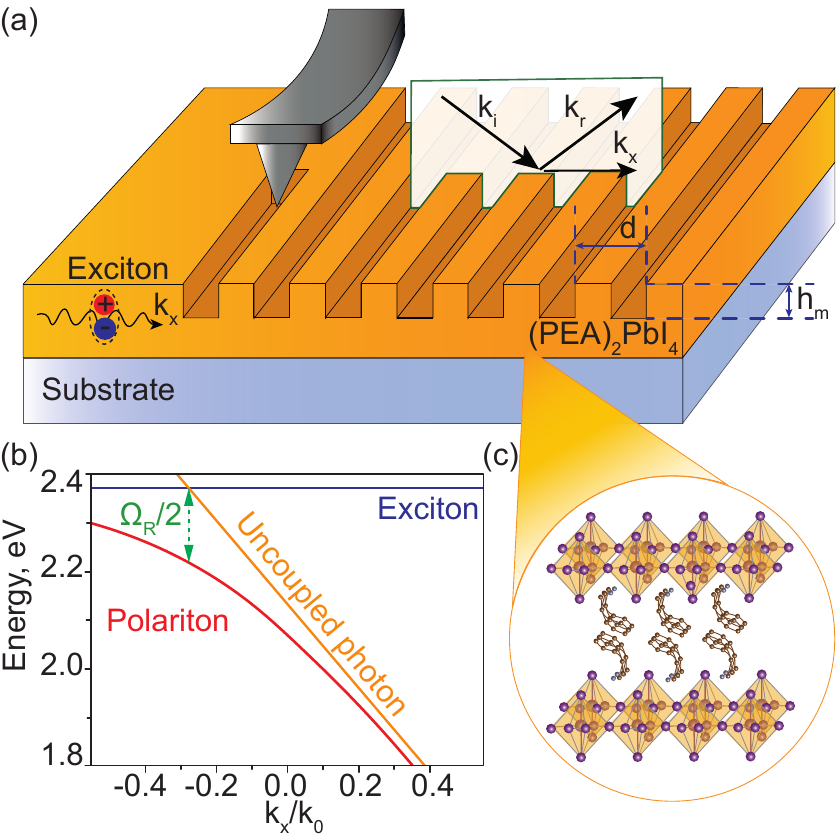}% Here is how to import EPS art
\caption{ (a) Sketch of mechanical scanning probe lithography of a thin $(\text{PEA})_2\text{PbI}_4$ film. AFM cantilever with a single-crystal diamond tip applies constant pressure on the film and moves with a highly precise trajectory to create a periodic structure of a photonic crystal slab (PCS). The inset schematically shows the incident ($\bm{k}_i$) and reflected ($\bm{k}_r$) wavevectors, as well as their in-plane component ($\bm{k}_x$). (b) Calculated dispersion of the lower polariton branch (red line) resulting from the strong coupling between the uncoupled exciton resonance (blue line) and the uncoupled photon cavity mode (orange line). (c) A sketch of the atomic structure of a 2D-perovskite  $(\text{PEA})_2\text{PbI}_4$.}
\label{fig:AFL}
\end{figure}

Previously, several methods for perovskite nanostructuring have already been demonstrated, however, all of them have disadvantages and limitations. The material degradation may be caused during focused ion beam and electron beam lithography \cite{alias2015focused,lamers2022perovskite}. Direct laser writing avoids this problem but has a limited lateral resolution above 200~nm.\cite{zhizhchenko2021directional} Nanoimprint method maintains the resolution of ion (or electron) beam lithography and does not cause degradation, but the stamp geometry can not be changed after its fabrication.\cite{masharin2022polaron,masharin2022room} From this point of view, mechanical scanning probe lithography (m-SPL)\cite{garcia2014advanced} (Fig. \ref{fig:AFL}a) appears one of the most versatile and convenient nanostructuring techniques for the perovskite planar exciton-polariton system since the mechanical cutting of perovskites does not cause material degradation, the atomic force microscopy (AFM) tip can be less than 10~nm in its lateral sizes, and high-precision piezo-stages of m-SPL allow for the dynamic tuning of various parameters of the resulting structure.

In this work, we demonstrate a universal and \red{affordable for research purposes} technology of 2D-perovskite film nanostructuring for the realization of room-temperature exciton-polariton planar cavities based on PCS with the precise control of polariton dispersion. By varying the period and modulation of PCS we change the exciton-polariton dispersion and its radiative lifetime. The developed m-SPL method for perovskites opens the way for the realization of planar polaritonic cavities with on-demand optical properties for nonlinear and active polaritonics.

%\section{Fabrication of planar perovskite cavities}
%\subsection{$(\text{PEA})_2\text{PbI}_4$ thin film synthesis}

\red{To fabricate the sample,} first, a thin film of 2D perovskite $(\text{PEA})_2\text{PbI}_4$ is synthesized by the solvent engineering method\cite{jeon2014solvent}. The solution for the synthesis is prepared by dissolving 149.4~mg of PEAI and 138.3~mg of PbI$_2$ in 1~ml of dimethylformamide. Before the synthesis of \red{the thin} film, we clean 12$\times$12~mm SiO$_2$ substrates with soapy water, acetone, and isopropanol consistently.\red{ To achieve the hydrophilic surface of the substrates, they are placed in the oxygen plasma cleaner for 10 minutes.} The synthesis of (PEA)$_2$PbI$_4$ films is performed in a glove box with a dry nitrogen atmosphere with the spin-coating method. \red{The prepared perovskite solution is deposited on top of the substrate and after it accelerates for 2 seconds and rotates at a speed of 4000 rpm for 60 seconds. The resulting film is annealed at 70$^{\circ}$ for 10 minutes.} The morphology of the synthesized film is studied with AFM, resulting in 130 nm film thickness and surface roughness of 15~nm (See Fig. S1 in Supplementary Information (SI)).

%Then it is necessary to evacuate the sample to slowly remove the solvent from the film and start slow crystallization, and then anneal it on a plate temperature $70^{\circ}$C in order to evaporate the solvent and form the desired perovskite phase. \textbf{you can not to comment, why you are doing this, if you cited correct work, where it is explained} The result was the  perovskite film with  thickness 130 nm and roughness 5 nm. \textbf{How did you know it?}

\begin{figure*}
\includegraphics[width=0.7\linewidth]{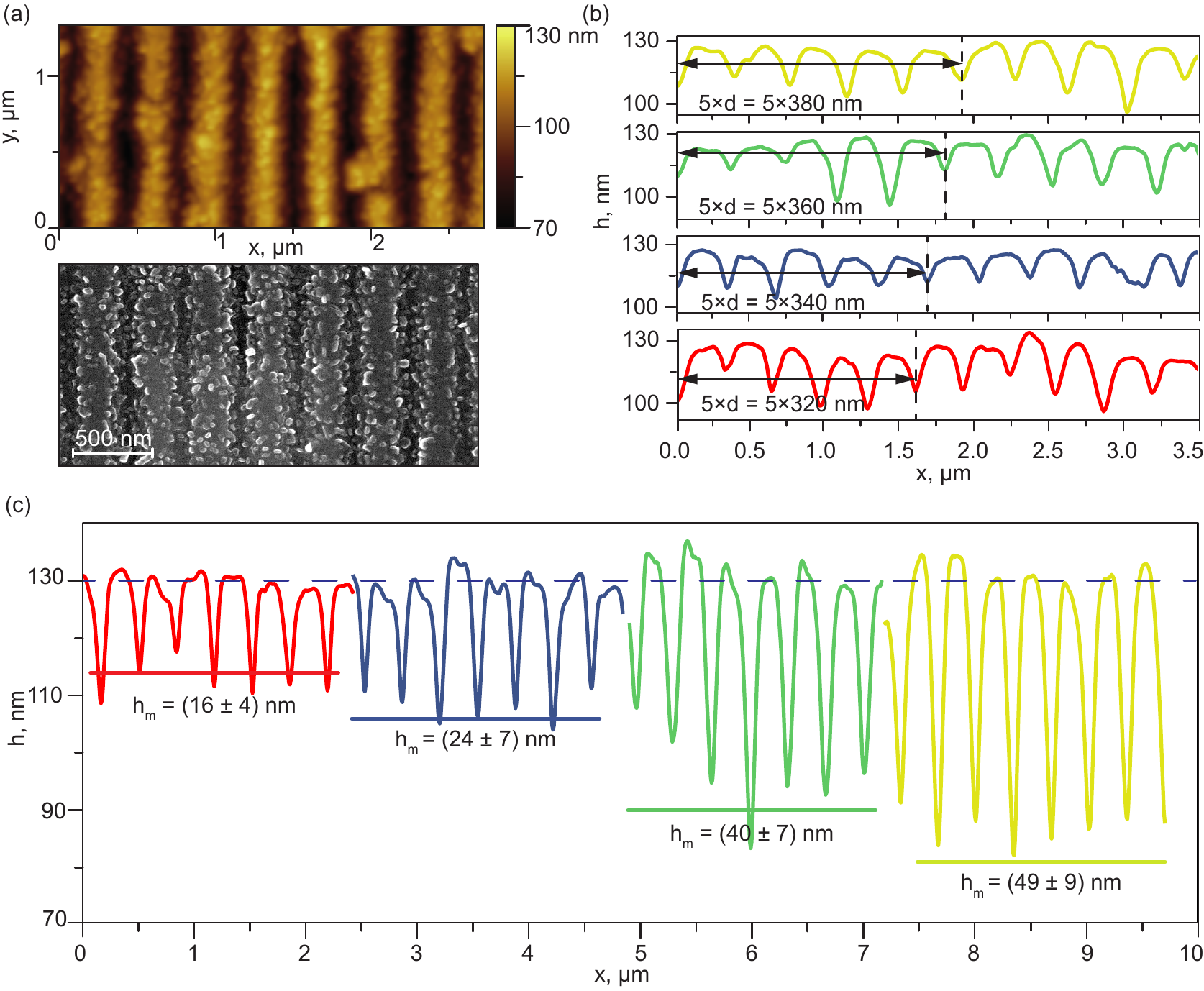}
\caption{\red{(a) The AFM \textcolor{black}{and (SEM)} images of a typical perovskite PCS fabricated with a period of \textcolor{black}{360}~nm. The pseudo color shows the topography of the PCS. (b) Extracted averaged profiles from AFM measurements of fabricated structures with various periods, the period changed within $d = 320 - 380$ nm, and the modulation is about $ h_m = 20$ nm.(c) Extracted averaged profiles of fabricated structures with various modulations, the modulation changes within $h_m= 16 - 49$ nm, and the period is equal to $d=340$ nm, "$\pm$" shows the standard deviation for each trench in the structure from the average modulation for the whole structure.}}
\label{fig:gratings}
\end{figure*}
\
%\subsection{Mechanical scanning probe lithography}
For m-SPL we use an atomic force microscope AIST-NT SMART SPM and cantilevers with a single-crystal diamond tip (TipsNano DRP-IN) with a resonant frequency of 500 – 1000 kHz, a normal spring constant of 350 N/m (See SI for the details), and a tip curvature radius of 25 – 35 nm. Before the lithography, the film morphology was characterized with AFM in a semi-contact regime. The use of the piezo-stages of the atomic force microscope makes it possible to control the diamond-tipped cantilever position with an accuracy of nanometers. Thus, fabricating the 1D PCS, we precisely control the period, the height modulation, and the trench width (see Fig.\ref{fig:AFL}a).

\red{During the fabrication, the cantilever is pushed towards the surface with force controlled by the AFM feedback system. The cantilever moves once along the specified direction, forming trenches in the perovskite film. During trench formation, the material is partially compressed and partially ejected towards the sides of the trench. However, most of it is moved by the tip to the end of the trench. At the end of each trench, the probe lifts off the film surface. The cut grains remain at the edge of the fabricated structure and do not affect the optical properties of the whole PCS.}   

The height modulation depends on the applied cantilever force, which is defined by the shift of the cantilever from the initial position and its stiffness. The force required to achieve a modulation of $15 - 50$ nm on $(\text{PEA})_2\text{PbI}_4$ perovskite films is experimentally \red{estimated} to be in the range of $5 - 30~ \mu\text{N}$. 
Since the tip is conical in shape, the minimum \red{trench} width depends on the modulation $h_m$. For modulation of $15 - 50$~nm, the width at the half-height of the \red{trench} is $80 - 130$ nm \red{(See SI for more details)}. 
The speed of the cantilever during the lithography process is limited \red{by} 3 $\mu \text{m}$/s because at higher values the probe begins to pull out perovskite grains. The optimal speed for the 2D-perovskite film lithography is found to be approximately 1 $\mu \text{m}$/s. 

By choosing the trajectory of the AFM tip with piezo-stages, this method allows the realization of mostly arbitrary structures. Particularly, it is possible to change the period of the PCS by programming the cantilever movement coordinates with nanometer precision. One of the most important advantages of m-SPL is the potential applicability of this method for the creation of PCSs, coupled waveguides, or other planar photonic designs on one 2D-perovskite film, combining them into one photonic on-chip system.

%\subsection{The fabricated photonic crystal slabs}

The fabricated PCSs have a lateral size of 15x30 $\mu\text{m}^2$. The typical morphology of the structure studied with AFM \textcolor{black}{and scanning electron microscopy (SEM)} is shown in Fig.~\ref{fig:gratings}a. By varying the cantilever displacement coordinate, we fabricated PCSs with the periods of $d=$ 320, 340, 360, 380 nm and modulation of about $h_m=20$~nm (Fig. \ref{fig:gratings}b). By changing the \red{pushing force} in the range of 9-24 $\mu\text{N}$, we also realize structures with different modulations of $h_m = 16,\ 24,\ 40,\ 49$~nm and a period of $d= 340$ nm (Fig. \ref{fig:gratings}c). Resulting structures are expected to have different spectral positions of the resonances and also different \red{coupling with the free space}, which we study further.

\begin{figure*}
\includegraphics[width=0.8\linewidth]{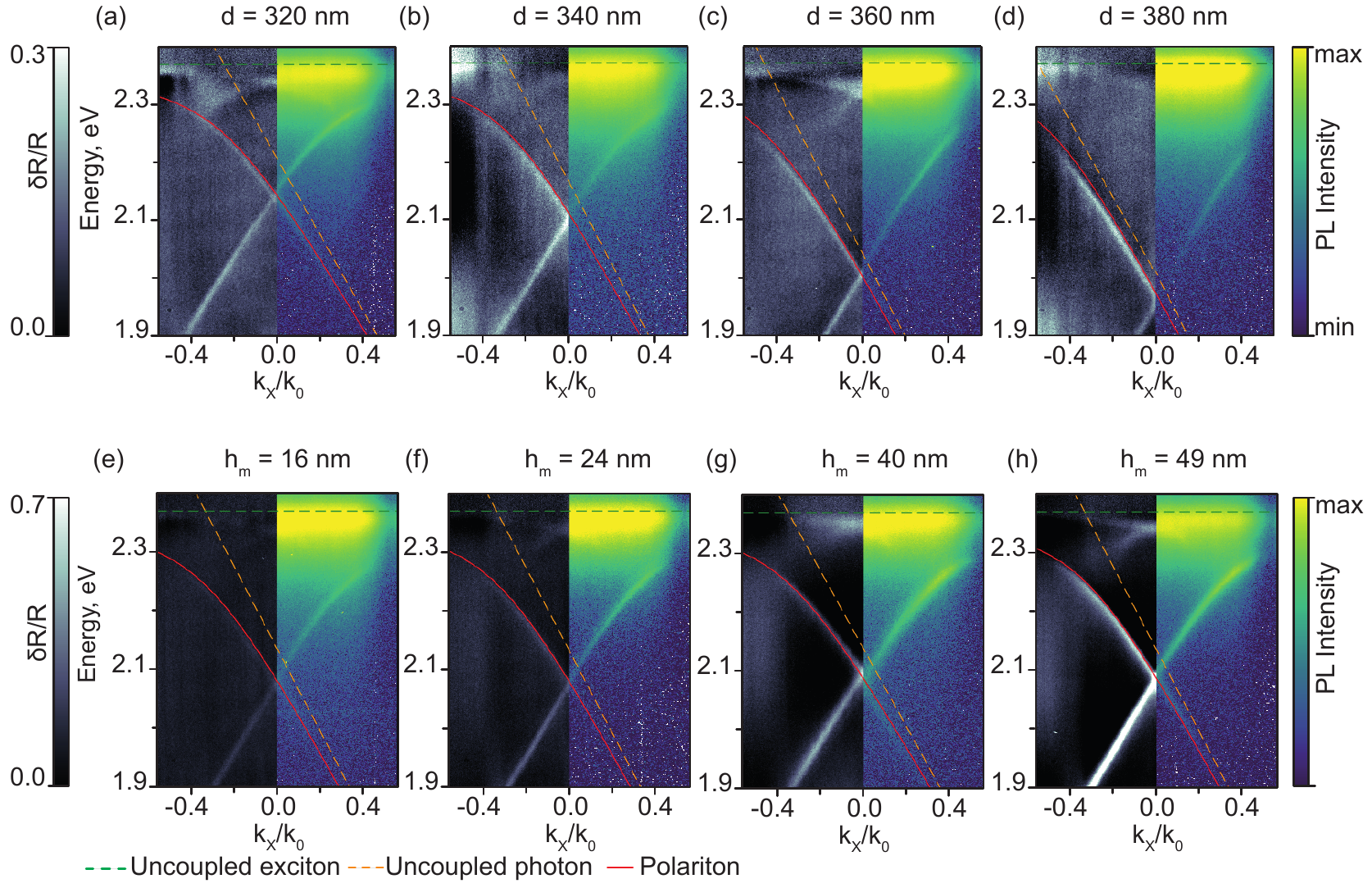}
\caption{ (a-d) Angle-resolved reflectance (left) and photoluminescence (right) spectra of the studied PCS based on the (PEA)$_2$PbI$_4$ with the periods $d = 320, 340, 360, 380$~nm, respectively, and the modulation $h_m \approx 20$ nm. (e-h) The same spectra maps of the fabricated PCSs with the period $d = 340$ nm and the modulations $h_m \approx 16, 24, 40, 49$~nm, respectively. Dashed yellow lines correspond to the estimated uncoupled photon cavity mode dispersions. Dashed green lines correspond to uncoupled exciton resonances. Red solid lines represent the polariton modes fitted with the two-coupled oscillators model.}
\label{fig:PLR}
\end{figure*}

%\section{Optical spectroscopy of the polaritons}
In order to study leaky \red{optical} modes of the fabricated PCSs, we perform angle-resolved spectroscopy measurements based on the back focal plane (BFP) setup. The BFP of the objective lens (Mitutoyo NIR $\times$50 with an N.A. of 0.55) is imaged on a slit spectrometer coupled to a liquid nitrogen-cooled imaging CCD camera (Princeton Instruments SP2500+PyLoN) using the 4f scheme (see SI for the details). For the illumination of the sample as well as for the measurements of the reflectance spectra, a halogen lamp is used. The plane of incidence contains both normal to the sample and the direction of periodicity of the PCS (see Fig.~\ref{fig:AFL}a). 
Before impinging the slit of the imaging spectrometer, light reflected from the sample passes through a linear polarizer aligned such that TE modes are studied. The scheme is also used to obtain the angle-resolved photoluminescence spectra using a femtosecond laser (Pharos, Light Conversion) coupled with a broad-bandwidth optical parametric amplifier (Orpheus-F, Light Conversion) at the wavelength of 480 nm, 100 kHz repetition rate as a non-resonant excitation source. All measurements were performed at a room temperature of 300K.  

We measure angle-resolved reflectance and photoluminescence spectra for every of the fabricated PCSs, shown in Fig.~\ref{fig:PLR}. The measured data show the pronounced leaky modes in the spectral region below the exciton resonance around 2.37~eV. All the studied samples demonstrate the curving of the mode dispersion asymptotically approaching the exciton level in the \red{high-frequency} region, revealing the signs of the strong light-matter coupling regime.\cite{masharin2022polaron} 

In order to verify the strong light-matter coupling regime, we extract the modes from the experimental data by the following procedure: first, we subtract the unbound exciton photoluminescence signal from the experimental dispersion at each $k_x/k_0$, then we fit the resulting modes by the peak Lorentz function data. Combining the spectral peak positions for each \red{in-plane wavevector} $k_x/k_0$, we obtain the extracted mode dispersion. Since the upper polariton branch (UPB) above the exciton resonance does not exist due to the strong non-radiative absorption, the only way to confirm the strong light-matter coupling regime is to fit the extracted mode with a lower polariton branch (LPB), estimated by the two-coupled oscillator model as\cite{hopfield1958theory}:

\begin{equation}
    E_{LP} = \frac{\widetilde{E}_x + \widetilde{E}_c(k)}{2} - \frac{1}{2}\sqrt{\left(\widetilde{E}_x - \widetilde{E}_c(k)\right)^2 +4 g^2 },
    \label{LPB}
\end{equation}
where $\widetilde{E}_x = E_x - i\gamma_x$ is complex energy accounting for the spectral position and the linewidth of the uncoupled exciton resonance, $\widetilde{E}_c(k) = E_c(k) - i\gamma_c$ is the complex dispersion of the uncoupled cavity photon mode, $g$ - is a light-matter coupling coefficient. The Rabi splitting $\Omega_R$ corresponds to the minimal energy distance between UPB and LPB, however, as UPB does not exist, we can only estimate this value based on the described model:
\begin{equation}
    \Omega_R = \sqrt{4g^2 - (\gamma_c - \gamma_x)^2};
    \label{eq:Rabi}
\end{equation}
%The strong light-matter coupling regime is confirmed, when the following criteria are satisfied :
%\begin{equation}
%    \begin{cases}
%        g \geq \frac{|\gamma_c - \gamma_x |}{2},\\ 
%        \Omega_R \geq \frac{\gamma_c + \gamma_x}{2}
%    \end{cases}
%    \label{eq:cond stong}
%\end{equation}

The uncoupled photon cavity mode has linear dependence of the energy on the wavenumber $k_x/k_0$ since the refractive index is considered to have negligible changes in the considered spectral range without accounting for the exciton resonance. Therefore, we estimate uncoupled photon cavity dispersions as $E_c(k_x) = k \times k_x + b$ based on the calculations of Fourier modal method\cite{li1997new} (see SI for details). The coupling coefficient $g$, as well as the half-widths of an unbound photon $\gamma_c$ and exciton $\gamma_x$, are chosen as the optimization parameters in the fitting of the LPB. The resulting real part of the PL dispersion $E_{LP}$ optimized for each of the samples is shown as red curves in Fig.~\ref{fig:PLR}. 

\begin{comment}
The exciton level $E_x$ was found from the reflection spectrum from the unstructured sample. The parameters of an uncouple photon $E_c(k_x) = k\cdot k_x + b$ were found as follows. In the case of structures with a change in the period, the tilt angle of an unbound photon $k$ was found from the calculated dispersion of the polariton by the Fourier modal method\cite{li1997new}. In this case, the free parameter $b$ was also found by the least squares method from the obtained experimental polariton dispersions. In the case of structures with modulation change, the parameters of an unbound photon $k, b$ were averaged for all structures and were fixed.
\end{comment}

\begin{figure}
\includegraphics{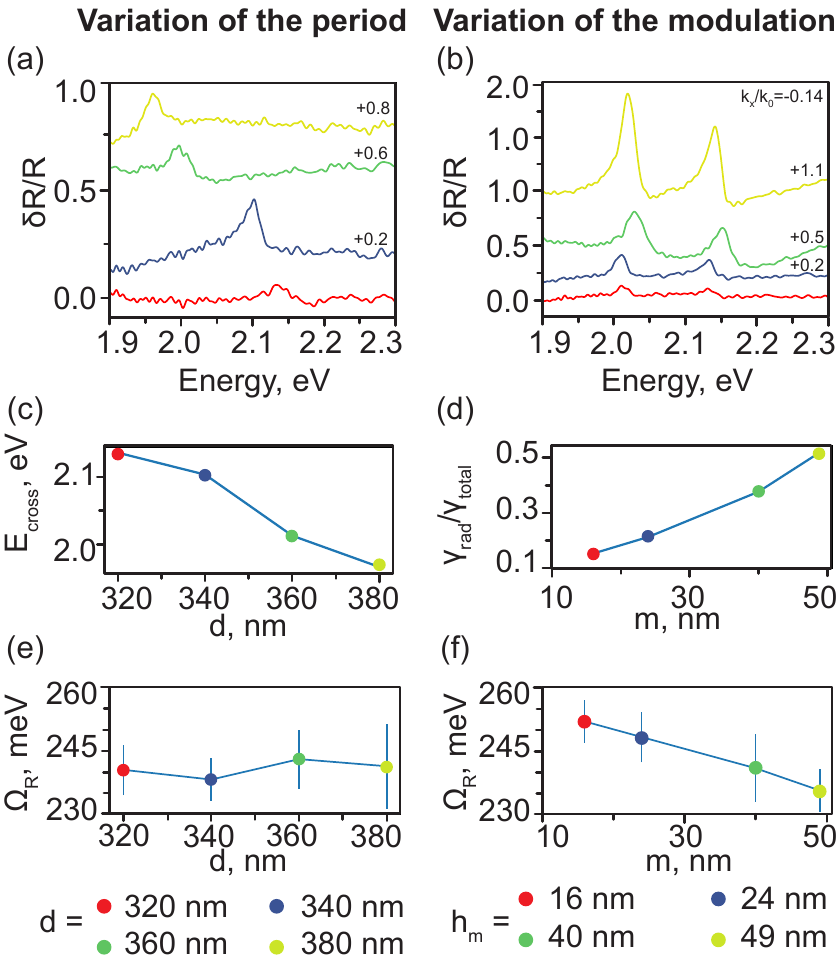}
\caption{(a) Reflectance spectra at $k_x/k_0 = 0$ for the \red{respective} PCS. They are shifted along the vertical axis at arbitrary values for better visual perception.\red {(b) Differentical reflectance spectra at $k_x/k_0 = 0.14$ for the fabricated PCSs. The spectra are shifted along the vertical axis at the specified values for better visual perception.}(c) The spectral position of the crossing point of counter-propagating polariton modes (at $\Gamma$-point) $E_{cross}$ depending on PCS period $d$. (d) The ratio between the radiative and total losses $\gamma_{rad}/\gamma_{total}$ depending on PCS modulation $h_m$. \red{On the charts (c) and (d) error bars do not exceed the size of the markers.} (e), (f) Rabi splitting $\Omega_R $ depending on PCS period $d$ and modulation $h_m$, respectively, \red{with error bars (See SI for details).} } 
\label{fig:4}
\end{figure}

The estimated values from the fitting of the uncoupled cavity photon $\gamma_c$ and exciton $\gamma_x$ do not exceed \textcolor{black}{23}~meV and  \textcolor{black}{18}~meV, respectively. The resulting values of Rabi splitting $\Omega_R$ for each of the PCSs are shown in Figs.~\ref{fig:4}e and \ref{fig:4}f, which \red{exceed 230 meV for all PCS modulations}. The obtained values satisfy the strong light-matter regime criteria  ($g>|\gamma_C-\gamma_X|/2$; $\Omega_R > |\gamma_C+\gamma_X|/2$)\cite{schneider2018two} \red{with a margin} in all studied samples. 
%Substituting the values into the condition (\ref{eq:cond stong}), we confirm the strong coupling of light and matter for all of PCSs. 

\

The leaky mode dispersion of the 1D PCS is determined by the waveguide modes folded towards the first Brillouin zone with the edges of $k^{BZ}_x = \pm \pi/d$, where $d$ is a PCS period. For a \textcolor{black}{planar} waveguide with the chosen thickness, with the change of the PCS period, the spectral position of folded uncoupled leaky modes, and, hence, polariton branches shift proportionally. \cite{gerace2004gap} Actually the difference in the spectral position of the polariton modes can be noticed in Figs.~\ref{fig:PLR}(a-d). In order to reveal the dependence of the spectral position of polariton mode as a function of period, we extract the reflection spectra at normal incidence $k_x/k_0 = 0$ (Fig.~\ref{fig:4}a). The frequencies of the  modes are estimated by fitting with the Fano resonance function\cite{miroshnichenko2010fano, fano1961effects, fan2003temporal} (See SI for details) and as expected show a monotonous decrease with the increase of PCS period (Fig.~\ref{fig:4}c). 

The value of the Rabi splitting $\Omega_R$ depends on the coupling coefficient and the linewidths of uncoupled exciton and cavity photon modes (Eq. \ref{eq:Rabi}). Since the coupling coefficient, $g$ depends on the cavity mode localization and oscillator strength\cite{hopfield1958theory,gambino2014exploring}, it should not change strongly with the PCS period or other geometrical parameters. This was confirmed by the results of fitting all the experimental data.  
In turn, the uncoupled exciton linewidth $\gamma_X$ is the property of the material and thus should not depend on the PCS design, which we also confirmed by analyzing the data. Hence, the only way to tune the Rabi splitting is to vary the radiative part of leaky mode losses $\gamma_C$, which is dictated by the PCS modulation. Thus, with the variation of the PCS period, we do neither expect nor observe the pronounced dependence of the estimated Rabi splitting $\Omega_R$ values (Fig.~\ref{fig:4}e).
 
The variation of the modulation $h_m$ with constant period provides \red{the change} in \red{differential reflection} contrast of the experimentally measured polariton modes, as shown in Figs.~\ref{fig:PLR}e-h and Fig.~\ref{fig:4}b. Higher modulation causes higher coupling of the leaky mode with the free space, or in other words, increases the radiative losses of the mode. In order to reveal the dependence, we estimate the ratio $\gamma_{rad}/\gamma_{total}$ by fitting the amplitude and asymmetry parameter of the Fano resonance (See SI for details) for different modulations $h_m$ at the $k_x/k_0$ = \red{0.14} and show it in Fig.~\ref{fig:4}d. Note that the non-radiative losses should be nearly constant for each of the PCSs because they are mostly dictated by the material defect states and excitonic absorption. Hence, the total optical losses $\gamma_C$ rise with increasing the PCS modulation, which leads to the \red{moderate} reduction of the Rabi splitting value (Fig.~\ref{fig:4}f). Thus by applying the different forces on the cantilever during the m-SPL process it is possible to control the modes contrast and the value of the Rabi splitting in the planar exciton-polariton PCS leaky modes.

\red{To conclude, in this work, we}  \textcolor{black}{have demonstrated} the method of mechanical scanning probe lithography for the realization of planar room-temperature exciton-polariton systems based on 2D perovskites. The fabricated PCSs \red{support} the long-living polariton modes with a Q-factor up to 100. Thanks to the \red{flexibility} of the m-SPL method, it is possible to vary the modulation of the structures with \red{a few tens of nanometers precision. The period of the PCS is controlled with nanometer precision. The width of the trenches is tens of nanometers.} In this way, we \textcolor{black}{have achieved full} control the dispersion, optical radiative losses, and the Rabi splitting of the exciton-polariton states in the planar photon cavity based on (PEA)$_2$PbI$_4$. Note that the demonstrated method can be introduced for other halide perovskites and for the fabrication of other planar photon cavities, including metasurfaces and PCSs. Our work reveals the \red{affordable for research purposes} and time-efficient method for the fabrication of planar high optical quality exciton-polariton systems based on 2D perovskite film, which is highly demanded for the realization of \textcolor{black}{non-equilibrium exciton-polariton condensation in perovskite metasurfaces \cite{Masharin2023Mar}} and optical nonlinear and active on-chip polaritonic devices.

\section*{Supplementary Information}
See supplementary information for more details on the surface morphology of the perovskite films, geometrical and mechanical properties of tip for m-SPL, description of the experimental setup, information on the processing of experimental data, fitting of polariton modes, and errors of the extracted qualities. 

\begin{acknowledgments}
\textcolor{black}{The authors thank Rumon Miah for the SEM characterization of the samples.} The experimental part of this work was funded by the Russian Science Foundation, grant \#21-12-00218 \textcolor{black}{and Ministry of Science and Higher Education of the Russian Federation (Project 075-15-2021-589)}. A.S. acknowledges the Deutsche Forschungsgemeinschaft (Grant SFB TRR142/project A6), the Mercur Foundation (Grant Pe-2019–0022), and TU Dortmund core funds.

\begin{comment}
We wish to acknowledge the support of the author community in using
REV\TeX{}, offering suggestions and encouragement, testing new versions,
\dots.
\end{comment}

\end{acknowledgments}

\nocite{*}
\bibliography{aipsamp}

% ****** Start of file aipsamp.tex ******
%
%   This file is part of the AIP files in the AIP distribution for REVTeX 4.
%   Version 4.1 of REVTeX, October 2009
%
%   Copyright (c) 2009 American Institute of Physics.
%
%   See the AIP README file for restrictions and more information.
%
% TeX'ing this file requires that you have AMS-LaTeX 2.0 installed
% as well as the rest of the prerequisites for REVTeX 4.1
% 
% It also requires running BibTeX. The commands are as follows:
%
%  1)  latex  aipsamp
%  2)  bibtex aipsamp
%  3)  latex  aipsamp
%  4)  latex  aipsamp
%
% Use this file as a source of example code for your aip document.
% Use the file aiptemplate.tex as a template for your document.
%\documentclass[%
% aip,
% jmp,
% bmf,
% sd,
% rsi,
% amsmath,amssymb,
%preprint,%
% reprint,%
%author-year,%
%author-numerical,%
% Conference Proceedings
%]{revtex4-1}

%\usepackage{graphicx}% Include figure files
%\usepackage{dcolumn}% Align table columns on decimal point
%\usepackage{bm}% bold math
%\usepackage[mathlines]{lineno}% Enable numbering of text and display math
%\linenumbers\relax % Commence numbering lines

%\usepackage[utf8]{inputenc}
%\usepackage[T1]{fontenc}
%\usepackage{mathptmx}
%\usepackage{etoolbox}
%\usepackage{comment}
%\usepackage{bm}
%\usepackage[usenames]{color}
%\usepackage{colortbl}
%% Apr 2021: AIP requests that the corresponding 
%% email to be moved after the affiliations

%\usepackage[normalem]{ulem}
%\usepackage{caption} %заголовки плавающих объектов
%\captionsetup[figure]{name=FIG. S}

%\usepackage[version=3]{mhchem} % Formula subscripts using \ce{}
\renewcommand{\thefigure}{S\arabic{figure}}
\renewcommand{\theequation}{S\arabic{equation}}
\renewcommand{\thesection}{S\arabic{section}}
\renewcommand{\thetable}{S\arabic{table}}

\makeatletter
\def\@email#1#2{%
 \endgroup
 \patchcmd{\titleblock@produce}
  {\frontmatter@RRAPformat}
  {\frontmatter@RRAPformat{\produce@RRAP{*#1\href{mailto:#2}{#2}}}\frontmatter@RRAPformat}
  {}{}
}%
\makeatother
%\begin{document}

%\preprint{AIP/123-QED}

\title[Mechanical scanning probe lithography of 2D perovskites for fabrication of planar polaritonic cavities]
{Supplementary Materials: Mechanical scanning probe lithography of perovskites for fabrication of high-Q planar polaritonic cavities}
% Force line breaks with \\
\author{N. Glebov}
  \altaffiliation{These authors contributed equally}
  %\altaffiliation{nikita.glebov@metalab.ifmo.ru}%Lines break automatically or can be forced with \\

\author{M. Masharin}%
\altaffiliation{These authors contributed equally}%Lines break automatically or can be forced with \\

\affiliation{ 
ITMO University, School of Physics and Engineering, St. Petersburg, 197101, Russia%\\This line break forced with \textbackslash\textbackslash
}%

\author{B. Borodin}%
\affiliation{ 
Ioffe Institute, Saint-Petersburg 194021, Russia%\\This line break forced with \textbackslash\textbackslash
}%
\author{P. Alekseev}%
\affiliation{ 
Ioffe Institute, Saint-Petersburg 194021, Russia%\\This line break forced with \textbackslash\textbackslash
}%

\author{F. Benimetskiy}%
\affiliation{ 
Department of Physics and Astronomy, University of Sheffield, S3 7RH, Sheffield, UK%\\This line break forced with \textbackslash\textbackslash
}%
\author{S. Makarov}%
\affiliation{ 
ITMO University, School of Physics and Engineering, St. Petersburg, 197101, Russia%\\This line break forced with \textbackslash\textbackslash
}%
\affiliation{ 
Qingdao Innovation and Development Center, Harbin Engineering University, Qingdao 266000, Shandong, China}%
\author{A. Samusev*}
 %\homepage{anton.samusev@gmail.com}
 \email{anton.samusev@gmail.com}
\affiliation{ 
ITMO University, School of Physics and Engineering, St. Petersburg, 197101, Russia%\\This line break forced with \textbackslash\textbackslash
}%
\affiliation{%
Experimentelle Physik 2, Technische Universit\"at Dortmund, 44227 Dortmund, Germany%\\This line break forced% with \\
}%
\date{\today}% It is always \today, today,
\maketitle
\onecolumngrid
%\appendix

\section{\label{sec:S1}Surface morphology of the pristine 2D perovskite film} %\protect\\ The line
\begin{figure*}[h!]
    \centering
    \includegraphics{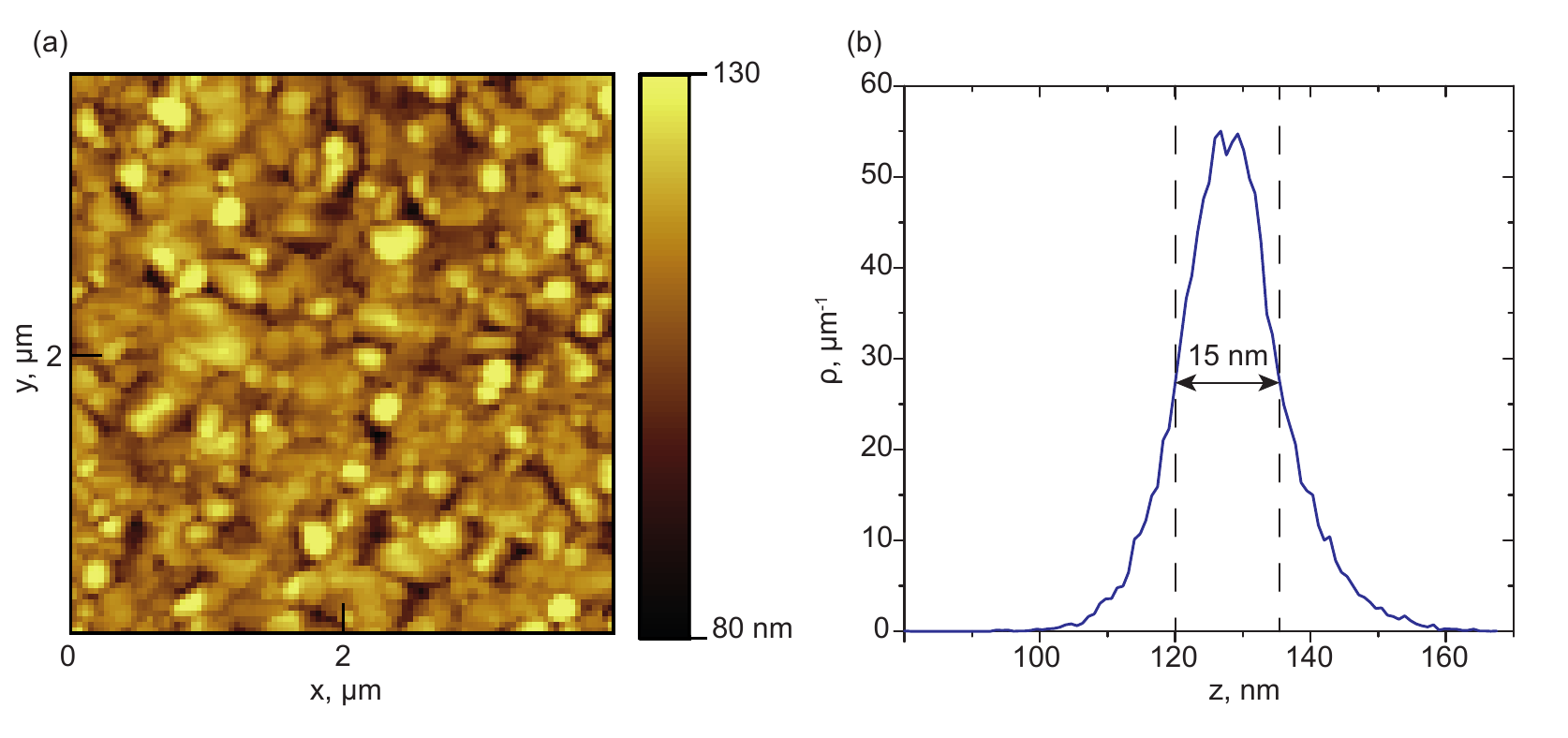}
    \caption{(a) AFM image of an unstructured sample $(\text{PEA})_2\text{PbI}_4$. (b) Height distribution of the sample determining the surface roughness.}
    \label{fig:S1}
\end{figure*}

%break was forced \lowercase{via} \textbackslash\textbackslash}
The topographical surface map of the pristine quasi-2D perovskite $(\text{PEA})_2\text{PbI}_4$ thin film is shown in Fig.~\ref{fig:S1}a. The topography is obtained by atomic force microscopy (AFM). Structuring by mechanical scanning probe lithography requires the roughness of the sample lower than the assumed modulation of the photonic crystal slab. The roughness can affect the non-uniform broadening of the optical modes, leading to a decrease in the radiative lifetime. Based on the obtained data we estimate the height distribution, shown in Fig.~\ref{fig:S1}b. The width at the half-height of this distribution determines the roughness of the sample. The roughness of the fabricated samples does not exceed 15 nm.  

\section{\label{sec:S2}Tip for mechanical scanning probe lithography}
\begin{figure}[h!]
    \centering
    \includegraphics[scale=0.8]{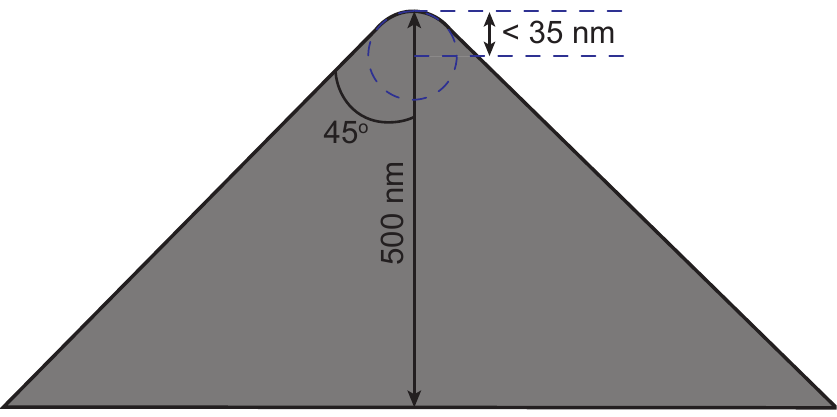}
    \caption{Sketch of the geometry of a single-crystal diamond tip used for mechanical scanning probe lithography.}
    \label{fig:S2}
\end{figure}

For the procedure of mechanical scanning probe lithography, we use an atomic force microscope SMART, cantilevers with a single-crystal diamond tip (TipsNano DRP-IN). The tip with a resonant frequency of 500 – 1000 kHz, a spring constant of 100 – 600 N/m, and a normal spring constant of 350 N/m. Normal spring constants for AFM tips used in experiments were calculated using the Sader method\cite{sader1999calibration}. This method allows us to determine the spring constant of an atomic force microscope cantilever using the following parameters: the resonant frequency, the quality factor of the cantilever in air, and their geometrical dimensions. The spring constant is given by:
\begin{equation}
     k=0.1906 \rho_{air}L b^2 Q \Gamma_i \omega _{vac}^{2}
     \label{eq:S1}
\end{equation}
where $\omega _{vac}$  is the resonant frequency of the cantilever in air, $b, L$ are width and length of the cantilever, $Q$ is the quality factor, $\rho _{air}$ is the density of the air and $\Gamma _i$ is the imaginary part of the so-called ‘‘hydrodynamic function’’\cite{green2005frequency}. This hydrodynamic function  $\Gamma(\omega)$ only depends on the Reynolds number $Re = \rho _{air} \omega b^2/(4 \eta)$, where $\eta$  is the viscosity of the surrounding environment and is independent of the cantilever thickness and density. We experimentally measured parameters such as $Q, L, b$ and $\omega$ for each of cantilever. The physical constants associated with the surrounding environment were taken from the literature:  $\rho _{air}= 1.85  \text{kg} \ m^{-3}$ and $\eta = 1.86 \times 10^{-5}  \text{kg} \ m^{-1}s^{-1}$.
Fig.~\ref{fig:S2} shows schematically the geometry of the tip used. The tip has a conical shape. The radius of curvature of its tip is less than 35 nm. The solution angle is 45 degrees. The height of the tip is ($500 \pm 100$) nm.  Thus, knowing the geometry of the probe used, it is possible to unambiguously determine the minimum dimensions of the cavities depending on the modulation depth. 

\section{\label{sec:S3}Pushing force}
For the procedure of mechanical scanning probe lithography, we set the pushing force on the perovskite thin film. It is possible to estimate the pushing force as\cite{mironov2004basics}: 
\begin{equation}
    F = k \Delta Z
    \label{eq:S2}
\end{equation}
where $k$ - the cantilever stiffness, $\Delta Z$ - the value characterizing the bending of the cantilever, induced by its shift with respect to the sample along the surface normal. In order to make the PCS with the necessary modulation, the pushing force was selected as follows. With some steps, starting with the minimum pressure for each following structure, we increased the force, until the huge pushing force completely carve the perovskite grains. We estimated the force required to obtain PCS with modulation in the 15-50 nm range based on quasi-two-dimensional perovskite $(\text{PEA})_2\text{PbI}_4$ using the probes indicated above for the m-SPL procedure.

\section{\label{sec:S4}Angle-resolved spectroscopy at room temperature}
\begin{figure*}[h!]
    \centering
    \includegraphics{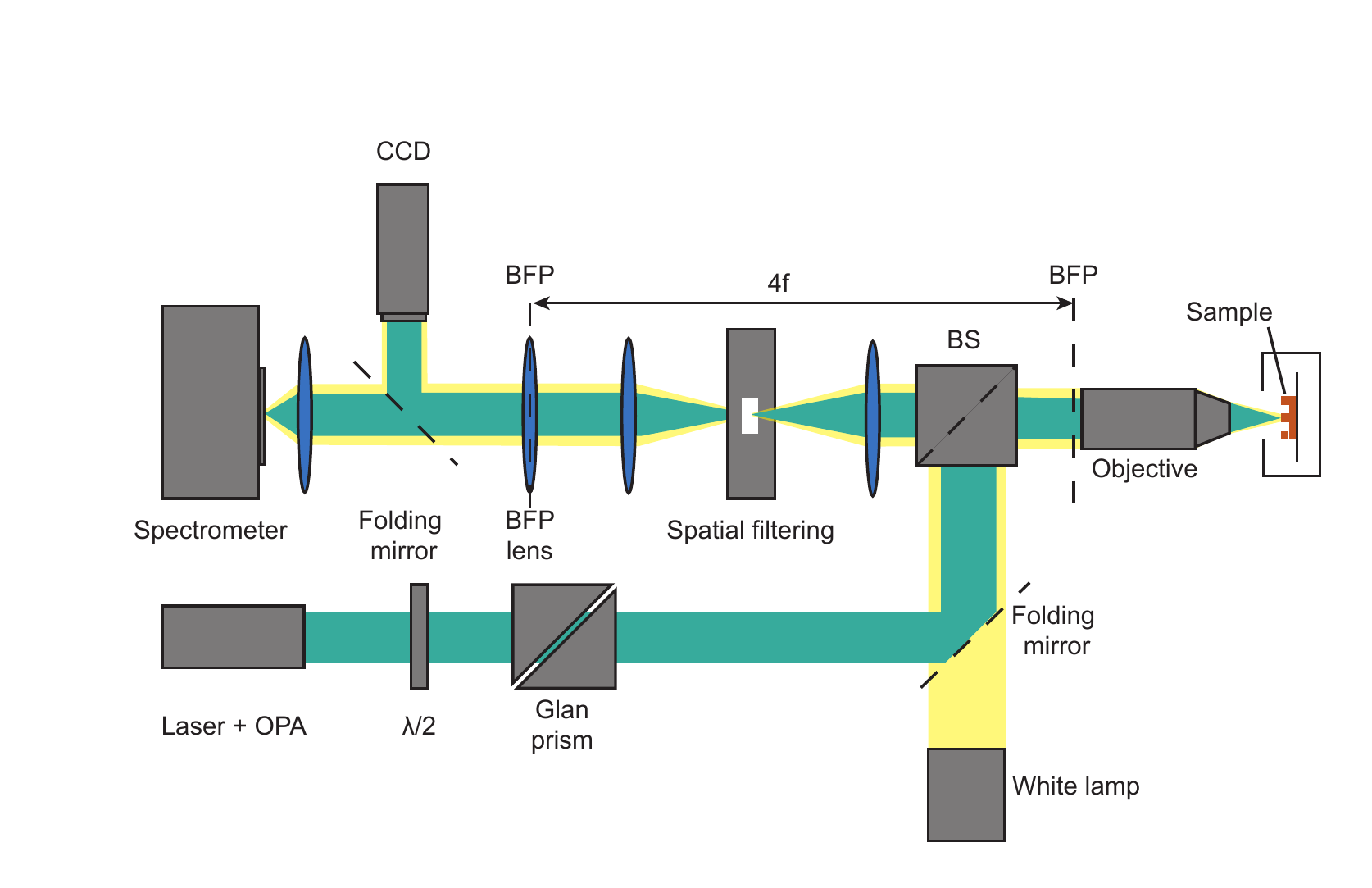}
    \caption{ The sketch of the experimental scheme of an experimental setup for angle-resolve reflectance and photoluminescence spectra measurements.}
    \label{fig:S3}
\end{figure*}

For the optical characterization of photonic crystal slabs (PCS), angle-resolved photoluminescence and reflection spectra were measured using the optical setup shown in Fig.~\ref{fig:S3}. The sample was placed in a chamber with a vacuum pump and optical window. The vacuum pump evacuates air up to pressure around 10$^{-5}$ bar to avoid the degradation of the perovskite sample from the oxygen. All measurements were carried out at room temperature. (See "optical experiment" section in the main text) 
\newpage
%A white halogen lamp serves as a source for measuring the optical response, while a femtosecond laser coupled to an optical parametric amplifier is used as a non-resonant pump in photoluminescence measurements.  Rear focal plane images were combined with real space filtering in a 4f configuration. This eliminates the unwanted background signal when obtaining spectra with angular resolution.  All signals were recorded using a Princeton Instruments spectrometer (f = 500 mm, 600 g/mm) and liquid nitrogen-cooled PyLoN eXcelon CCD camera.

\section{\label{sec:S5}Spectral processing of experimental data}
\begin{figure*}[h!]
    \centering
    \includegraphics{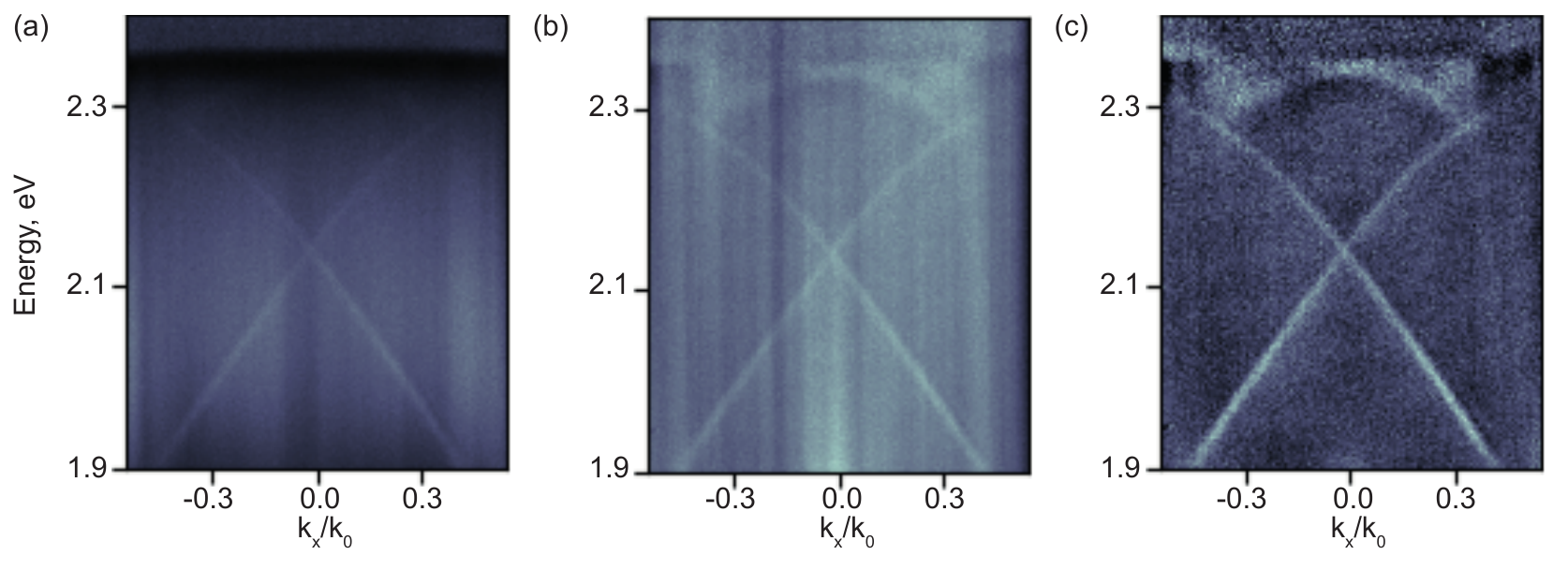}
    \caption{(a) Angle-resolved reflection spectrum from PCS based thin film  $(\text{PEA})_2\text{PbI}_4$ in the absence of spectral processing. (b) Angle-resolved reflection spectrum normalized to wavelength averaged values at each of the angles. (c) Angle-resolved reflection spectrum using averaging from (b) and normalized data to the straight line obtained from the detrend.}
    \label{fig:S4}
\end{figure*}

In order to make the angle-resolved reflection spectra more contrast, we perform the data treatment shown in Fig.\ref{fig:S4}. The first step is to normalize the reflection spectra at each wavenumber $\bm{k}_x/\bm{k}_0$ by the average value of the spectrum in terms of energy at that wavenumbers $\bm{k}_x/\bm{k}_0$ (Fig.\ref{fig:S4}b). The second step is to find a linear function which is an underlying background of the spectra at each wavenumber $\bm{k}_x/\bm{k}_0$. Then the spectra at each wavenumber $\bm{k}_x/\bm{k}_0$ are normalized to the corresponding linear functions and the final data are obtained (Fig.\ref{fig:S4}c).

\section{\label{sec:S6}Fitting of polariton dispersions}
\begin{figure*}[h!]
    \centering
    \includegraphics{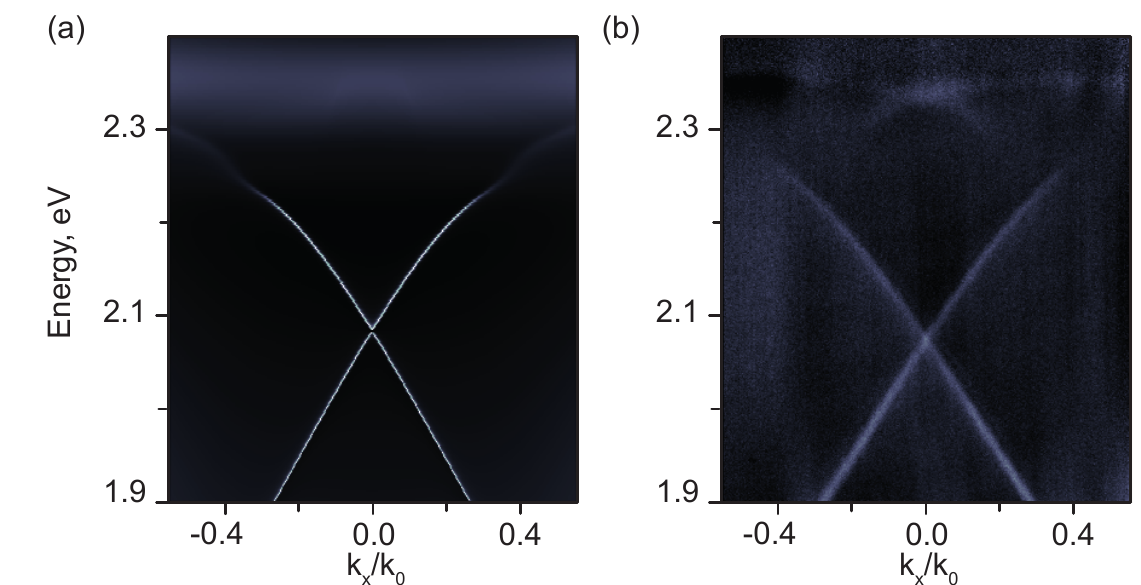}
    \caption{(a) Angle-resolved reflection spectrum of a structured $(\text{PEA})_2\text{PbI}_4$ sample calculated by the Fourier modal method. (b) Angle-resolved reflection spectrum of a structured $(\text{PEA})_2\text{PbI}_4$ measured in experiment.}
    \label{fig:S5}
\end{figure*}

To extract the polariton mode from the angle-resolved photoluminescence spectrum, we first subtract the uncoupled exciton resonance signal from the measured spectrum at each wavenumber $\bm{k}_x/\bm{k}_0$ and fit the data by a Lorentz peak function. Combining the positions of the peaks at all wavenumbers $\bm{k}_x/\bm{k}_0$, we obtain the polariton dispersion.  It is possible to estimate the coupling strength between cavity photon and exciton resonance using the fitted bound oscillator \cite{hopfield1958theory}, a model that takes into account the spectral position and linewidth of the unbound exciton $\widetilde{E}_x = E_x - i\gamma_x$ and the photon mode parameters of the unbound cavity  $\widetilde{E}_c(k) = E_c(k) - i\gamma_c$. Polariton states are defined as:
\begin{equation}
    E_{LP} = \frac{\widetilde{E}_x + \widetilde{E}_c(k)}{2} - \frac{1}{2}\sqrt{\left(\widetilde{E}_x - \widetilde{E}_c(k)\right)^2 +4 g^2 }
    \label{eq:S3}
\end{equation}
where $g$ – is the light-matter coupling strength. To use this model for parameter fitting, it is necessary to estimate the parameters of the uncoupled excitation resonance and photon. The exciton level $E_x$ was found from the reflection spectrum from the unstructured sample. The uncoupled photon parameters $E_c(k_x) = k\times k_x + b$ for structures with different periods or modulations were determined in different ways. For a period-changing PCS, the slope coefficient of an unbound photon $k$ is determined from the calculated dispersion of a polariton by the modal Fourier method \cite{li1997new} (See Fig.\ref{fig:S5}a). 
The slope coefficient was estimated from the extrapolation of the polariton dispersion to large $k_x$. The free parameter $b$ was estimated by the least squares method from fitting the data of experimental polariton dispersions. For a PCS with a change in modulation, the parameters of an unbound photon $k, b$ were estimated from the modal Fourier method. Then were the fitting parameters in the model of two coupled oscillators. After that they were averaged and fixed for all structures.

\section{\label{sec:S7}Fitting polariton modes with the Fano profile}
\subsection{\label{sec:S71} Fano resonance in optical cavities}
Using the theoretical model of a single-mode optical resonator associated with 2 ports, one can obtain the intensity reflection coefficient:\cite{fan2003temporal}
\begin{equation}
    R = \frac{r^2(\omega - \omega_0)^2 + t^2(1/\tau)^2 \pm 2rt(\omega - \omega_0)(1/\tau)}{(\omega - \omega_0)^2 + (1/\tau)^2}
    \label{eq:S4}
\end{equation}
where $\omega_0$ and $\tau$ are the center frequency and the lifetime
of the resonance, respectively. $r$ and $t$ are real parameters of S-matrix with $r^2+t^2 = 1$. Sign $\pm$ corresponds to the parity of the resonant mode with respect to the mirror plane. In all systems except those with r or t being zero, the spectral profile of the reflectance coefficient has Fano line shape. 

Let us reduce the formula (\ref{eq:S4}) to the form in which it is used to fit the experimental data. Total losses of the system are $\gamma_{total} = 1/\tau$. Using the formula (\ref{eq:S4}) we obtain:
\begin{equation}
    R = \frac{r^2\left( \frac{\omega - \omega_0}{\gamma_{total}} \pm \frac{\sqrt{1-r^2}}{r}\right)^2}{\left(\frac{\omega - \omega_0}{\gamma_{total}}\right)^2 +1}
    \label{eq:S5}
\end{equation}
Let's change the parameters  $r^2 = A$, $\frac{\omega - \omega_0}{\gamma_{total}} = \varepsilon$, $\pm \frac{\sqrt{1-r^2}}{r} = q$ and $B = A q^2$. Thus we obtained a formula (\ref{eq:S5}) for fitting optical modes in reflection. 

\subsection{\label{sec:S71}Fit Fano function with losses}
\begin{figure*}[h!]
    \centering
    \includegraphics[width=0.85\linewidth]{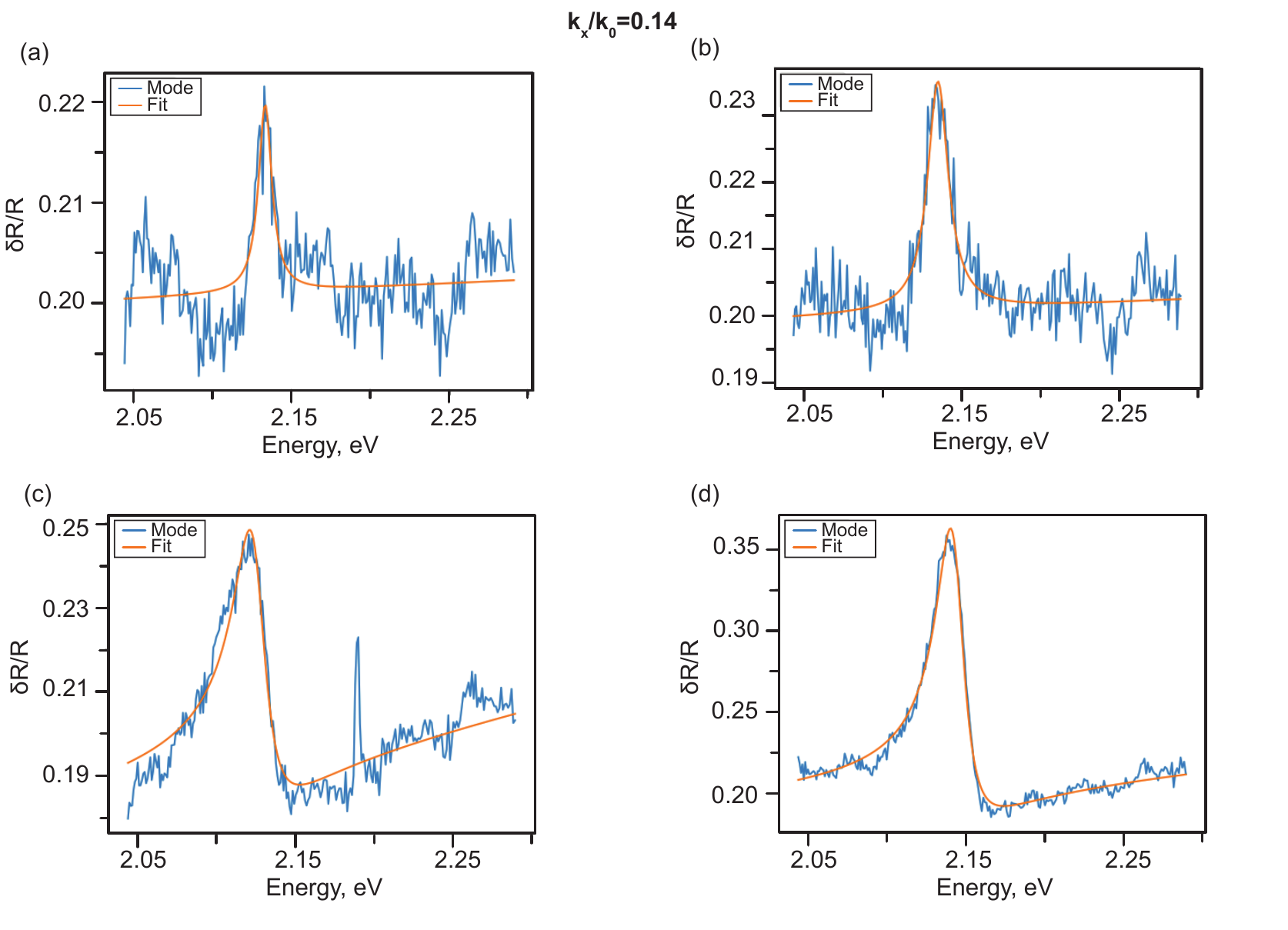}
    \caption{(a)-(d) Fitting the Fano function of the angle-resolved reflection spectrum at a fixed angle $k_x/k_0 = 0.14$ for PCS with modulation $h_m = 16, 24, 40, 49~\text{nm}$, respectively}
    \label{fig:S6}
\end{figure*}

The Fano function with an amplitude coefficient was used to fit the parameters of the exciton-polariton modes in reflection, shown in Fig.~\ref{fig:S6}\cite{fano1961effects}:
\begin{equation}
    f_F(\varepsilon) = B \frac{(\frac{\varepsilon}{q} + 1)^2}{\varepsilon^2 + 1}
    \label{eq:S6}
\end{equation}
where B is the amplitude coefficient, q is the phenomenological shape parameter, $\varepsilon = (E - E_c)/\gamma_{tot}$.  $E_c$ is the resonant
energy and  $\gamma_{tot}$ is the width of the autoionized state. In the limit $|q| \to \infty $, the shape of the line is determined by the transition through a discrete state with a Lorentz profile, as the transition to the continuum is very weak \cite{miroshnichenko2010fano}. The total loss can be represented as a sum of radiative and nonradiative losses $\gamma_{tot} = \gamma_{rad} + \gamma_{nrad}$, then the Lorentz profile can be expressed as\cite{ivchenko2005optical} : 
\begin{equation}
    f_L = \frac{\gamma_{rad}^2}{(E-E_c)^2 + (\gamma_{rad} + \gamma_{nrad})^2}
    \label{eq:S7}
\end{equation}
The parameters $B,\ q, \ E_c,\ \gamma_{tot}$ are estimated during the fit by the Fano function of the experimental reflectance spectrum at $\bm{k}_x/\bm{k}_0 = 0.14$. We use the limit pass of the Fano resonance to the Lorentz profile to express the ratio of the radiative and total losses.
\begin{equation}
     f_F \underset{|q| \gg \varepsilon}=f_L 
     \label{eq:S8}
\end{equation}
\begin{equation}
     f_F \underset{|q| \gg \varepsilon}=  \frac{B}{\varepsilon^2 + 1} = B \frac{ (\gamma_{rad} + \gamma_{nrad})^2}{(E-E_c)^2 + (\gamma_{rad} + \gamma_{nrad})^2} =f_L 
     \label{eq:S9}
\end{equation}
Thus it is possible to relate the fit parameters to the ratio of radiative and total losses: 
\begin{equation}
    \frac{\gamma_{rad}}{\gamma_{rad} + \gamma_{nrad}} = \sqrt{B}
    \label{eq:S10}
\end{equation}

\newpage
\section{\label{sec:S8}Mechanical scanning probe lithography of $(\text{PEA})_2\text{PbI}_4$ and $\text{MA}\text{PbBr}_3$}

\begin{figure*}[h!]
    \centering
    \includegraphics[width=0.9\linewidth]{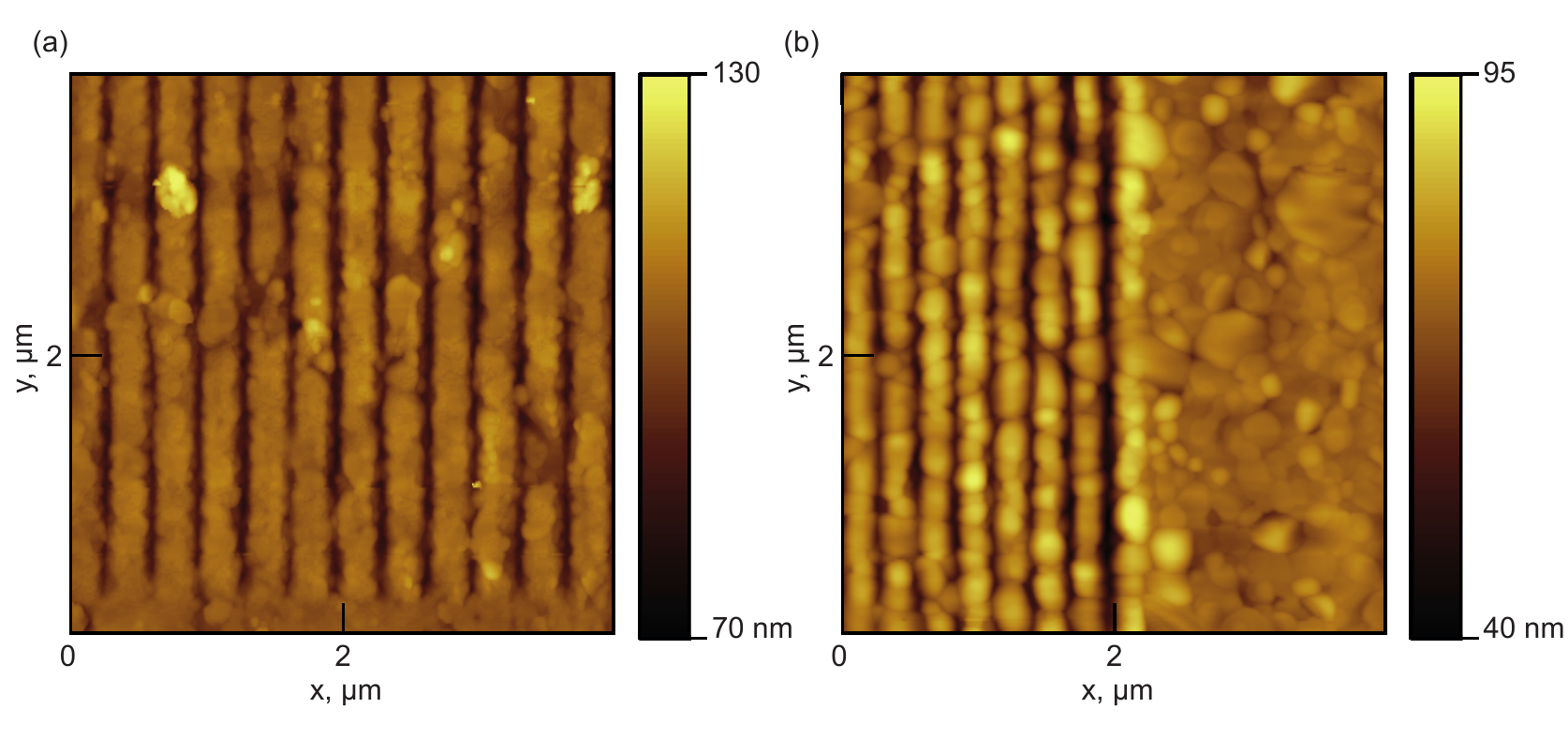}
    \caption{(a) The AFM image of the fabricated $(\text{PEA})_2\text{PbI}_4$ PCS with a period of 340 nm and modulation of 49 nm. (b) The AFM image of the fabricated PCS based on the halide perovskite $\text{MA}\text{PbBr}_3$ with a period of 280 nm and modulation of 45 nm.}
    \label{fig:S7}
\end{figure*}
Mechanical scanning probe lithography is an excellent method for structuring other perovskites as well, including 3D perovskites. As an example, the structuring of a thin film of $\text{MA}\text{PbBr}_3$ perovskite was carried out (See Fig.~\ref{fig:S7}b).

\section{\label{sec:S9}Errors of the extracted quantities}

\begin{table*}[h]
\caption{\label{tab:S1} Table of estimated Rabi splitting values, intersection point energy, and ratio of radiation to total loss with their errors.}
\begin{ruledtabular}
\begin{tabular}{ccccccccc}
 &\multicolumn{4}{c}{Period}&\multicolumn{4}{c}{Modulation}\\
 \  &320 nm &340 nm& 360 nm & 380 nm & 16 nm& 24 nm & 40 nm & 49 nm\\ \hline
 $\Omega_{Rabi}$, meV  & 240 &  238 & 243 & 241 & 252 & 248 & 240 & 235 \\
 RMS error, meV &   6 &   5 & 7 &  10 &  5 & 6 & 10 & 5 \\  \hline
 $E_{cross}$, eV&2.14 & 2.10 & 2.01 & 1.97 \\ 
  RMS error, eV& $3 \times 10^{-6}$ & $1 \times 10^{-6}$& $0 \times 10^{-7}$ & $2 \times 10^{-6}$\\ \hline
 $\gamma_{rad}/\gamma_{total}$& & & & & 0.14 & 0.19 & 0.24 & 0.40\\
RMS error & & & & & $1.7 \times 10^{-3}$ & $1.5 \times 10^{-3}$ & $1.8 \times 10^{-3}$ & $1.8 \times 10^{-3}$ \\
\end{tabular}
\end{ruledtabular}
\end{table*}

%\nocite{*}
%\bibliography{aipsampsuppl}

%\end{document}

\end{document}